\documentclass[%
 reprint,
superscriptaddress,
 amsmath,amssymb,
 aps,
]{revtex4-2}

\usepackage{graphicx}
\usepackage{dcolumn}
\usepackage{bm}

\usepackage{subcaption}

\usepackage[dvipsnames]{xcolor}

\begin{document}

\preprint{APS/123-QED}

\title{Coarse-Grained Force Fields via Rotational Entropy Corrections to Free Energy Landscapes of Diffusing Molecules}

\author{J. M. Hall}
\email{jhall4@uoregon.edu, jmonroe.hall4@gmail.com}

 \affiliation{%
 Department of Physics, University of Oregon, Eugene, Oregon 97403, USA}%

\author{M. G. Guenza}%
 \email{mguenza@uoregon.edu}

\affiliation{%
 Department of Chemistry and Biochemistry, Department of Physics, Material Science Institute, and Institute for Fundamental Science, University of Oregon, Eugene, Oregon 97403, USA
}%

\date{\today}

\begin{abstract}
The construction of accurate interatomic potentials, and related fields of forces, from equilibrium conformational distributions of molecules is a crucial step in coarse-grained modeling. In this work we show that in order to develop accurate lab-frame force fields that preserve translational and rotational diffusion of a molecule, the observed body-fixed free energy landscape must be corrected for conformation-dependent rotational entropy to isolate the potential energy surface. We further demonstrate that even when the instantaneous effects of the correction are small, the resulting lagged correlations of the modeled force can be greatly altered and hence the correction is especially vital when parameterizing friction coefficients using modeled interatomic potentials.
\end{abstract}

\maketitle

Coarse-grained (CG) models of solvated molecules often rely on the potential of mean force (PMF) to capture the equilibrium conformational distributions. When the gradient of the PMF is used, it also defines the corresponding force field governing the system’s dynamics \cite{Note}. The PMF is often obtained via the Boltzmann factor from the distribution of molecular configurations observed in computer simulations \cite{Roux1995, Trzesniak2007}. While progress has been made on techniques for optimizing the potential for use in Langevin Equations to simulate reaction coordinates, for example through machine learning, much of this work has been focused on low-dimensional models of purely internal degrees of freedom describing the shape of the molecule. Such models do not include the translational and rotational diffusion of the molecule or any possible rovibrational coupling \cite{Lange2008a,Straub1990,Best2009,Best2010, leeDarveMultidimAlanine}. In many biological applications, such as protein–ligand binding, the timescales of slow conformational fluctuations and overall orientational dynamics can be comparable, making it desirable to explicitly model all relevant degrees of freedom for a comprehensive understanding of the process. In this work, we present a formalism to calculate an accurate lab-frame force field with the appropriate rot+trans symmetries by modifying the PMF obtained via inversion of the Boltzmann factor from the distribution of the shape coordinates.

A coarse-grained model of a molecule composed of $N$ point particles has $3N$ total degrees of freedom, $\vec{r}$. However, because the potential energy surface is invariant under global translations and rotations, it depends only on a reduced set of $3N-6$ internal shape coordinates, $\vec{q}$. These coordinates can be defined by removing the effect of translation and rotation through appropriate frame constraints \cite{Chapuisat1981}. Despite this reduction, the six global degrees of freedom remain implicitly present in the coarse-grained space and can couple to the internal dynamics, introducing a residual rotational entropy in the free energy landscape over $\vec{q}$. This entropy contribution is not associated with the internal potential $V(\vec{q})$. Therefore, to obtain a force field corresponding solely to $V(\vec{q})$ by inverting the Boltzmann distribution, the shape-dependent rotational entropy must be properly accounted for and removed to isolate the contribution of the potential energy surface.

If we were interested only in a body-fixed coarse-grained description, for example, of a single internal reaction coordinate, then the translational and rotational degrees of freedom would be eliminated through the projection procedure itself, with their influence absorbed into the memory kernel and noise. It is our decision to keep these symmetries explicit in our CG model that forces us to grapple with their effects on the free energy landscape. This choice, while more complex, enables us to capture global diffusion and its coupling to internal dynamics, providing a foundation for coarse-grained models of processes such as molecular binding.

\section{Body-Fixed Coordinates}

A freely diffusing molecule may be modeled as $N$ point particles, which can represent atoms or the centers of coarse-grained clusters, for a total of $3N$ spatial coordinates. 

The general Hamiltonian for the $N$ particles in Cartesian coordinates is given by

\begin{equation}
    \begin{aligned}
        H = \frac{1}{2}\vec{p}M^{-1}\vec{p}^{T}+V(\vec{r}).\\
    \end{aligned}
\end{equation}

Here M is a mass matrix consisting of a $3\times3$ diagonal block $m_{i}I_{3}$ for each particle $i$. The vector $\vec{p}$ consists of the xyz-momenta of each of the $N$ particles such that $\vec{p} = (p_{1x}, p_{1y}, p_{1z},...,p_{Nx},p_{Ny},p_{Nz})$, and likewise $\vec{r}$ represents the lab-frame positions. The potential energy $V(\vec{r})$ will have translational and rotational symmetry and so can be completely described by a reduced set of $3N-6$ shape coordinates. 

As in \cite{Chapuisat1981}, the six global coordinates corresponding to the overall translation and rotation can be defined via constraint equations on the $3N$ configuration coordinates. The choice of constraints defines the body-fixed frame and fixes the overall position $\vec{R}$ and Euler angles $\vec{\Theta}$ given a configuration $\vec{r}$. An example set of constraints are the Eckart conditions \cite{Eckart1935}, which are linear equations and identify $\vec{R}$ with the center-of-mass of the $N$ particles while defining the $\vec{\Theta}$ such that rotational-vibrational coupling in vacuum is minimized near a reference configuration. Once the constraints have been chosen, a reduced set of $3N-6$ internal shape coordinates $\vec{q}$ may also be chosen to complete the body-fixed description. Again following the approach in \cite{Chapuisat1981}, the Hamiltonian may be recast, for any choice of constraints, in canonical body-fixed coordinates as

\begin{equation}
    H = \frac{1}{2}\vec{P}M^{-1}\vec{P}^T + \frac{1}{2}\vec{p}S^{-1}\vec{p}^T + \frac{1}{2}\vec{\mathcal{J}}I^{*^{-1}}\vec{\mathcal{J}}^T + V(\vec{q}). \label{eq:bodyfixed_Hamiltonian}
\end{equation}

The first term is the kinetic energy associated with the center-of-mass, where $\vec{P}$ is the total momentum of the molecule and $M$ the total mass. The second term is the kinetic energy associated with the vibrational motion of the shape coordinates $\vec{q}$, which we have the freedom to choose, and $S$ is a mass-weighted metric defined as

\begin{equation}
S_{ij} := \sum_{k,\alpha} m_k \frac{\partial r_{k\alpha}}{\partial q_i}\frac{\partial r_{k\alpha}}{\partial q_j}. \label{eq:S_matrix}
\end{equation}

(Throughout this manuscript, Greek indices denote Cartesian components.) The third term represents the kinetic energy associated with a generalized angular momentum $\vec{\mathcal{J}}$, corresponding to the total angular momentum less the contribution from vibrations in the body-fixed frame, where

\begin{equation}
    \begin{aligned}
        \vec{\mathcal{J}} :&= I^{*}\vec{\omega} \ ,\label{generalizedAngularMomentum}
    \end{aligned}
\end{equation}

and

\begin{equation}
    \begin{aligned}
        I^{*} := I - CS^{-1}C^{T} \, \label{generalizedInertia}
    \end{aligned}
\end{equation}

with

\begin{equation}
C_{\alpha i} :=  \sum_{k} m_k (\vec{r}_k \times \frac{\partial \vec{r}_k}{\partial q_i})_{\alpha}. \label{coriolisMatrix}
\end{equation}

The generalized angular momentum may also be expressed entirely in canonical coordinates via

\begin{equation}
    \begin{aligned}
        \vec{\mathcal{J}} &=B(\vec{\Theta})\vec{p}_{\Theta}-CS^{-1}\vec{p},\label{generalizedAngularMomentum_canonical}
    \end{aligned}
\end{equation}
where $\vec{p}$ and $\vec{p}_{\Theta}$ are the canonical momenta associated with $\vec{q}$ and $\vec{\Theta}$ respectively, and

\begin{equation}
    B(\vec{\Theta}) := \begin{pmatrix} sin(\chi) & -csc(\theta)cos(\chi) & cot(\theta)cos(\chi) \\ cos(\chi) & csc(\theta)sin(\chi) & -cot(\theta)sin(\chi) \\ 0 & 0 & 1 \end{pmatrix}
\end{equation}
as in the treatment of rovibrational coupling in \cite{wilsonDeciusCross}.

We are free to choose a set of internal coordinates to describe the shape of the molecule. Certain coordinate systems will have inherent advantages and disadvantages. The Bond-Angle-Torsion (BAT) coordinates are popular and have the advantage of being rotation-invariant, and hence calculable in any frame\cite{Chang2003,Chang2006,Hikiri2016}. They are fundamentally curvilinear coordinates, however, and we will see that their state-dependent metric, via the $S$ matrix (\ref{eq:S_matrix}), introduces other difficulties. We will instead, in this work, accept the inconvenience of frame-switching and work with a reduced set of rectilinear displacements from a body-fixed reference structure (see Supplemental Material \cite{supplementalMaterial}).

To obtain the distribution of internal displacements, the translational and rotational motion of the molecule must be removed from the lab-frame trajectory. This procedure is inherently ambiguous, so appropriate frame constraints must be applied. To eliminate translational motion, we fix the center of mass at the origin. To remove rotational motion, we adopt the Eckart frame constraints. \cite{Eckart1935,Louck1976}. This choice is again motivated by the desire to maintain linearity in the constraint equations, which ensures a constant metric tensor. In contrast, using the quadratic constraints of the principal axes frame would result in a curvilinear metric for the internal coordinates.

\section{Rotational Entropy Correction to the Force Field in a Coarse-Grained Representation}

If the atomistic system is Boltzmann-distributed, then its probability distribution in phase-space is

\begin{equation}
    \begin{aligned}
        \mathcal{P}^{atom}(\vec{r}, \vec{p})d\vec{r}\:d\vec{p}&=
        \frac{1}{Z}e^{
        -\big(\frac{1}{2}\sum\limits_{i}\frac{1}{m_{i}}p_{i}^{2}+V^{atom}(\vec{r})\big)/k_{B}T}d\vec{r}\:d\vec{p} \\
    \end{aligned}
\end{equation}
where $\vec{r}, \vec{p}$ are the lab-frame positions and momenta of the atoms. The Boltzmann distribution doesn't strictly apply in the case of unbounded translational diffusion, but the simulated system can be made normalizable by a finite box or periodic boundary conditions and the influence of these constraints made sufficiently small as desired.

We will consider the case where the desired CG variables are the lab-frame centers-of-mass $\vec{r}^{L}$ of designated clusters of atoms in the molecule, and their corresponding momenta $\vec{p}^{L}$. We note that the remaining degrees of freedom in each cluster can be represented by the relative Jacobi coordinates $\vec{r}^{J}, \vec{p}^{J}$ while preserving the decoupled quadratic form of the kinetic energy. This allows us to trivially integrate out these degrees of freedom, together with those of the solvent and other ignored atoms, 
yielding the distribution in the CG phase space:
\begin{equation}
    \begin{aligned}
        \mathcal{P}^{L}(\vec{r}^{L}, \vec{p}^{L})d\vec{r}^{L}\:d\vec{p}^{L}&=
        \frac{1}{Z}e^{
        -\big(\frac{1}{2}\sum\limits_{n}\frac{1}{m_{n}}p^{L^{2}}_{n}+V(\vec{r}^{L})\big)/k_{B}T}d\vec{r}^{L}\:d\vec{p}^{L}\ , \ \label{eq:cg_labdist}
    \end{aligned}
\end{equation}
where
\begin{equation}
    \begin{aligned}
        V(\vec{r}^{L})&:=-k_{B}T\hspace{1mm}ln\bigg(\int d\vec{r}_{orth}\hspace{1mm}e^{-V^{atom}(\vec{r})/k_{B}T}\bigg). \label{eq:pmf}
    \end{aligned}
\end{equation}
Owing to translational and rotational invariance, the potential of mean force $V$ for a freely diffusing molecule depends only on the $3N-6$ shape coordinates $\vec{q}$, i.e., $V = V(\vec{q})$.

Our objective is to derive an expression for the observable probability distribution $\mathcal{P}(\vec{q})$ in terms of the potential of mean force $V(\vec{q})$ by integrating out the center-of-mass and rotational (Euler) degrees of freedom from \ref{eq:cg_labdist}. 

To this end, we begin by performing a canonical transformation from the lab-frame coordinates to a body-fixed representation, following the approach derived in \cite{Chapuisat1981} and summarized in Section I.

\begin{equation}
(\vec{r^{L}}, \vec{p^{L}}) \rightarrow (\vec{q}, \vec{p}, \vec{R}, \vec{P}, \vec{\Theta}, \vec{p}_{\Theta})
\end{equation}
where $\vec{q}, \vec{p}$ are the internal shape coordinates and their conjugate momenta, $\vec{R}, \vec{P}$ are the center-of-mass of the CG sites and its momentum, $\vec{\Theta}=(\theta,\phi,\chi)$ are the Euler angles, and $\vec{p}_{\Theta}= (p_{\theta}, p_{\phi}, p_{\chi})$ are the conjugate momenta to the Euler angles.

Since the transformation is canonical, Liouville's theorem implies that the phase space volume is preserved, and thus the probability distribution retains a simple volume element:

\begin{equation}
    \begin{aligned}
        d\vec{r}^{L}\:d\vec{p}^{L} \rightarrow d\vec{R}\:d\vec{P}\:d\vec{\Theta}\:d\vec{p}_{\Theta}\:d\vec{q}\:d\vec{p}.\\
    \end{aligned}
\end{equation}

Again following \cite{Chapuisat1981}, the Hamiltonian is now expressed in the body-fixed coordinates by Eq. (\ref{eq:bodyfixed_Hamiltonian}).


The Hamiltonian's independence from $\vec{R}$ implies a uniform distribution over center-of-mass coordinates, which may thus be trivially integrated out.  Moreover, the center-of-mass momentum decouples from the remaining degrees of freedom and can likewise be integrated out, yielding:

\begin{equation}
    \begin{aligned}
        \mathcal{P}^{0}(\vec{\Theta}, \vec{p}_{\Theta}, \vec{q}, \vec{p})&=
        \frac{1}{Z}e^{
        -\big(\frac{1}{2}\vec{p}S^{-1}\vec{p}^T + \frac{1}{2}\vec{\mathcal{J}}I^{*^{-1}}\vec{\mathcal{J}}^T + V(\vec{q})\big)/k_{B}T} \ .\ \label{eq:cg_labdist}
    \end{aligned}
\end{equation}

A Gaussian integration over the internal shape momenta yields:

\begin{equation}
    \begin{aligned}
        \int d\vec{p}\:e^{-\big(\frac{1}{2}\vec{p}S^{-1}\vec{p}^{T}\big)/k_{B}T} = k_{B}T\sqrt{(2\pi)^{3N-6}}\sqrt{(|S(\vec{q})|)} \ .
    \end{aligned}
\end{equation}

After incorporating the constants into the normalization, the expression becomes:

\begin{equation}
    \begin{aligned}
        \mathcal{P}^{1}(\vec{\Theta}, \vec{p}_{\Theta}, \vec{q})&=
        \frac{1}{Z}\sqrt{|S(\vec{q})|}e^{
        -\big(\frac{1}{2}\vec{\mathcal{J}}I^{*^{-1}}\vec{\mathcal{J}}^T + V(\vec{q})\big)/k_{B}T}\ . \ \label{eq:cg_labdist}
    \end{aligned}
\end{equation}

Due to the simultaneous dependence of $\vec{\mathcal{J}}(\vec{\Theta}, \vec{p}_{\Theta})$ on $\vec{\Theta}$ and $\vec{p}_{\Theta}$, there is no simple way to integrate out these coordinates. Instead, we perform a coordinate transformation from the canonical momenta $\vec{p}_{\Theta}$ to the angular velocity $\vec{\omega}$. Rewriting the first term of the exponent in terms of $\vec{\omega}$ gives us

\begin{equation}
    \begin{aligned}
        \frac{1}{2}\vec{\mathcal{J}}I^{*^{-1}}\vec{\mathcal{J}}^T &= \frac{1}{2}\vec{\omega}I^{*}\vec{\omega}^{T} \ .\\ \label{eq:cg_labdist}
    \end{aligned}
\end{equation}

Since the transformation is noncanonical, care must be taken with the volume element, which transforms according to the determinant of the Jacobian as

\begin{equation}
    \begin{aligned}
        d\vec{p}_{\Theta} &= |J|d\vec{\omega} \ ,\\ \label{eq:cg_labdist}
    \end{aligned}
\end{equation}
where
\begin{equation}
    \begin{aligned}
        J &= \frac{\partial \vec{p}_{\Theta}}{\partial \vec{\omega}}.\\ \label{eq:cg_labdist}
    \end{aligned}
\end{equation}

We find this derivative by inverting Eq. (\ref{generalizedAngularMomentum_canonical}) to express $\vec{p}_{\Theta}$ in terms of the remaining coordinates and $\vec{\omega}$:

\begin{equation}
    \begin{aligned}
        \vec{p}_{\vec{\Theta}}&=B^{-1}(\vec{\Theta})\left(I^{*}(\vec{q})\vec{\omega}+C(\vec{q})S^{-1}(\vec{q})\vec{p}\right).\\
    \end{aligned}
\end{equation}

It follows that

\begin{equation}
    \begin{aligned}
        \frac{\partial \vec{p}_{\Theta}}{\partial \vec{\omega}}&=B^{-1}(\vec{\Theta})I^{*}(\vec{q}) \ ,\\
    \end{aligned}
\end{equation}
and the Jacobian determinant is 
\begin{equation}
    \begin{aligned}
        |J|&=\bigg|\frac{\partial \vec{p}_{\Theta}}{\partial \vec{\omega}}\bigg|\\
        &=\bigg|B^{-1}(\vec{\Theta})I^{*}(\vec{q})\bigg|\\
        &=\bigg|B^{-1}(\vec{\Theta})\bigg|\bigg|I^{*}(\vec{q})\bigg|.\\
    \end{aligned}
\end{equation}

Upon evaluating the determinant of $B^{-1}(\vec{\Theta})$, we find 

\begin{equation}
    \begin{aligned}
        |J|&=sin(\theta)\bigg|I^{*}(\vec{q})\bigg|.\\
    \end{aligned}
\end{equation}

Thus, after the coordinate transformation, the probability distribution becomes

\begin{equation}
    \begin{aligned}
        \mathcal{P}(\vec{\Theta}, \vec{\omega}, \vec{q})&=
        \frac{1}{Z} sin(\theta) |I^{*}(\vec{q})| \sqrt{|S(\vec{q})|}\:e^{
        -\big(\frac{1}{2}\vec{\omega}I^{*}\vec{\omega}^T + V(\vec{q})\big)/k_{B}T}.\\ \label{eq:cg_labdist}
    \end{aligned}
\end{equation}

The dependence on $\phi$ and $\chi$ has dropped out, and the $sin(\theta)$ term integrates to a constant, leaving

\begin{equation}
    \begin{aligned}
        \mathcal{P}(\vec{\omega}, \vec{q})&=
        \frac{1}{Z} |I^{*}(\vec{q})| \sqrt{|S(\vec{q})|}\:e^{
        -\big(\frac{1}{2}\vec{\omega}I^{*}\vec{\omega}^T + V(\vec{q})\big)/k_{B}T}.\\ \label{eq:cg_labdist}
    \end{aligned}
\end{equation}

Finally, performing the Gaussian integral over $\vec{\omega}$ yields the final expression for the probability distribution:

\begin{equation}
    \begin{aligned}
        \mathcal{P}(\vec{q})&=
        \frac{1}{Z} \sqrt{|I^{*}(\vec{q})|} \sqrt{|S(\vec{q})|}\:e^{
        -V(\vec{q})/k_{B}T}.\\ \label{eq:cg_labdist}
    \end{aligned}
\end{equation}

We find that the resulting distribution of shape coordinates, expressed via their potential of mean force, deviates from what a naive application of the Boltzmann factor would predict. The landscape is modified by the $\vec{q}$-dependence of both $I^{*}$ and $S$. Notably, this result holds for arbitrary choices of shape coordinates and frame constraints.

The shape-dependence of $S$ is merely a geometric artifact introduced by the use of curvilinear shape coordinates and/or nonlinear frame constraints. As discussed in the Supplemental Material, this dependence can be removed by an appropriate choice of coordinates and constraints. For the remainder of this work, we assume such a choice has been made, rendering $S$ a constant that may be absorbed into the normalization.

By contrast, the shape-dependence of $I^{*}$  is an inherent consequence of the rotational entropy associated with a given conformation and must be evaluated.

The connection to rotational entropy becomes more apparent upon rewriting the probability distribution as

\begin{equation}
    \begin{aligned}
        \mathcal{P}(\vec{q})&=
        \frac{1}{Z} \:e^{
        -\big(V(\vec{q})-TS_{R}(\vec{q}))\big)/k_{B}T},\\ \label{eq:cg_labdist}
    \end{aligned}
\end{equation}
where the rotational entropy $S_{R}$ associated with the coarse-grained shape $\vec{q}$ is defined by
\begin{equation}
    \begin{aligned}
        S_{R}(\vec{q})&:=
        k_{B}\:ln(A\sqrt{|I^{*}(\vec{q})|}),\\ \label{eq:cg_labdist}
    \end{aligned}
\end{equation}
and
\begin{equation}
A:= \frac{\pi^{1/2}}{\sigma}\left(\frac{8\pi^{2} e k_{B} T}{h^{2}}\right)^{3/2},
\end{equation}
is a normalization constant. The symmetry number $\sigma$ accounts for the symmetry of the reference conformation, taking values greater than one for molecules with rotational self-symmetry, and unity otherwise \cite{McQuarrie2000}.

For a rigid body, $I^{*}$  reduces to a constant matrix equal to the standard moment of inertia, yielding the familiar rigid-body expression for the rotational entropy of the reference conformation $\vec{q}=0$.\cite{McQuarrie2000} For flexible molecules, the conformation-dependent moment of inertia must be replaced by the generalized tensor of Eq.\ref{generalizedInertia}. Accordingly, the potential of mean force $V(\vec{q})$ relates to the probability distribution $\mathcal{P}(\vec{q})$ as follows:

\begin{equation}
    \begin{aligned}
        V(\vec{q})&=
        -k_{B}T\:ln(\mathcal{P}(\vec{q})) + k_{B}T\:ln(A\sqrt{|I^{*}(\vec{q})|}).\\ \label{eq:cg_labdist}
    \end{aligned}
\end{equation}

The generalized force conjugate to the $i$th shape coordinate is therefore

\begin{equation}
    \begin{aligned}
        -\frac{\partial V(\vec{q})}{\partial q_{i}}&=k_{B}T\:\frac{1}{\mathcal{P}(\vec{q})}\frac{\partial \mathcal{P}(\vec{q})}{\partial q_{i}} - \frac{k_{B}T}{2}\:\frac{1}{\big|I^{*}(\vec{q})\big|}\frac{\partial}{\partial q_{i}}\big|I^{*}(\vec{q})\big|.\\ \label{eq:cg_labdist}
    \end{aligned}
\end{equation}

The first term is obtained from an analytical fit to the probability distribution, as described below. During numerical calculation, it may be useful to replace the derivative of the determinant in the second term with the derivative of a matrix using Jacobi's formula:

\begin{equation}
    \begin{aligned}
        -\frac{\partial V(\vec{q})}{\partial q_{i}}&=k_{B}T\:\frac{1}{\mathcal{P}(\vec{q})}\frac{\partial \mathcal{P}(\vec{q})}{\partial q_{i}} - \frac{k_{B}T}{2}Tr\big(I^{*^{-1}}(\vec{q})\frac{\partial I^{*}(\vec{q})}{\partial q_{i}}\big).\\ \label{eq:force field}
    \end{aligned}
\end{equation}

\section{Parameterization of the Force Field for a Model Trimer}
\label{forcefield}
\begin{figure}[t]
\includegraphics[width = 8.6cm]{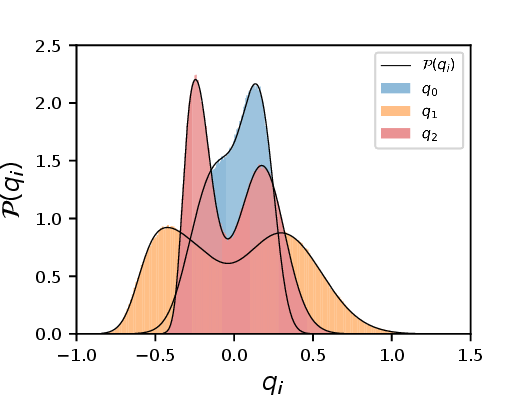}
\caption{\label{fig:pdf_fit} Histograms (transparent) of the shape coordinate distributions obtained from simulations of the model trimer, overlaid with the corresponding marginal distributions predicted by a 10-component Gaussian Mixture Model (GMM) fit (solid lines).}
\end{figure}

Representative samples from the distribution $\mathcal{P}(\vec{q})$ can be obtained from long, ergodic simulations of a molecule, for example through Molecular Dynamics (MD) simulations of a protein in solvent. To study the coarse-grained dynamics of that molecule, one needs to define the CG sites by grouping together a number of atoms. Then, by integrating out the fast degrees of freedom using a projection operator  expressed in the space of the coarse-grained coordinates, one derives a Generalized Langevin Equation (GLE) that guides the CG dynamics of the molecule. \cite{Mori1965, Zwanzig2001}

Over sufficiently long time intervals, where the protein dynamics is uncorrelated with the solvent dynamics, the GLE reduces to a Markovian Langevin equation in the position, ${\vec{r}}_{i}(t)$, and velocity, $\vec{v}_{i}(t)=\vec{p}_{i}(t)/m_i$,  of each CG site, $i$. The resulting Langevin equation,

\begin{eqnarray}
\dot{\vec{r}}_{i}(t)=&&\frac{1}{m_{i}}\vec{p}_{i}(t), \nonumber
\\
\dot{\vec{p}}_{i}(t)=&& -\frac{\partial V(\vec{r}(t))}{\partial \vec{r}_{i}}-\sum_{j}\zeta_{ij}\vec{v}_{j}(t) +\vec{f}^{R}_{i}(t), \nonumber
\\
\label{eq:langevin}
\end{eqnarray}
describes the time evolution of the CG variables in the field of the fast variables projected out. The eliminated degrees of freedom affect the dynamics through the form of the effective potential $V(\vec{r})$, the random forces $\vec{f}^{R}_{i}$, and the friction coefficients $\zeta_{ij}$.

The friction coefficients are obtained from the time integral of the memory kernel. In recent work, we demonstrated that these coefficients can be computed as a ratio of time correlation functions, highlighting that an accurate representation of the friction requires an equally accurate force-field parameterization \cite{frictionHallGuenza}. Consequently, to rigorously evaluate the accuracy and dynamical impact of the rotational entropy correction in coarse-grained (CG) models, it is essential to employ a system where all quantities entering the Langevin equation are known exactly.

As a test system, we consider a model trimer composed of three beads with masses $m = (3,4,3)$ and Markovian friction coefficients $\zeta = (10, 10, 20)$. All the simulation quantities are in reduced units \cite{allen_computer_1987,helfand_principle_1960}.
The internal potential governing  the trimer conformation is given by
\begin{equation}
\begin{aligned}
U(l_{1},l_{2},\theta) &= \frac{k_{1}}{2}(l_{1}-l_{1_{0}})^{2} + \frac{k_{2}}{2}(l_{2}-l_{2_{0}})^{2}\\
&+ \frac{k_{\theta}}{2}[(\theta-\theta_{0})^{2}(\theta-(\pi-\theta_{0}))^{2}-b(\theta-\frac{\pi}{2})^{2}], \label{eq:toy_potential}
\end{aligned}
\end{equation}
where $l_{1}, l_{2}$ are the bond lengths and $\theta$ is the angle between them. The trimer thus has a double well in the bond angle symmetric around $\theta=\pi/2$. The well locations are defined by $(l_{1_{0}}, l_{2_{0}}, \theta_{0}) = (1,1,\pi/3)$. The other well parameters are chosen as $(k_{1}, k_{2}, k_{\theta}, b) = (40, 40, 28, 1.5)$.

The stochastic noise is modeled as Gaussian-distributed, consistent with the Central Limit Theorem, which justifies this approximation in real systems where the random force arises from the cumulative effect of numerous solvent collisions. The temperature is set to $k_{B}T = 5$ via by the second fluctuation-dissipation theorem \cite{VanKampen2007}:

\begin{equation}
\langle f^{R}_{i\alpha}(t) f^{R}_{j\beta}(t')\rangle = 2k_{B}T\zeta_{ij}\delta(t-t')\delta_{\alpha\beta}. \label{fluctuationDissipation}
\end{equation}

Using this model system, we investigate the influence and quantitative impact of the rotational entropy correction on both the reconstruction of the full force field and the calculation of friction coefficients. We simulate Equation (\ref{eq:langevin}) using the Euler-Maruyama method with time step $\Delta t = 0.01$ for a total time of $T_{sim} = 10^{4}$.

To calculate the force field, we need to transform the trajectory into body-centered dynamics. Thus, 
after simulating Eq. (\ref{eq:langevin}) in the  laboratory frame, we derive the shape coordinates in the corresponding Eckart frame. We construct a reference structure $\vec{r}_{0}$ by aligning the configuration of one of the wells,  $(l_{1}=1, l_{2}=1, \theta=\pi/3)$, into a principal axes frame. Using this reference, we then compute the time series of Eckart-frame Euler angles from the laboratory-frame trajectory via the quaternion-based method described in \cite{Krasnoshchekov2014}. The shape coordinates $\vec{q}$ are defined as the first $3N-6$ elements of $P(\vec{r}_{B} - \vec{r}_{0})$, where the $\vec{r}_{B}$ are the body-fixed positions of the beads and $P$ is the permutation matrix

\begin{equation}
P_{ij}=
\begin{cases}
\delta_{ij},& i \not\in (3N-6,\:3N-5)\\
\delta_{i+1,j}& i = 3N-6\\
\delta_{i-1,j}& i = 3N-5\\
\end{cases}
\end{equation}

With this permutation matrix the last bead $N$ is omitted from the shape coordinates entirely, bead $(N-1)$ retains only an $x$-coordinate in the body-fixed frame, and bead $(N-2)$ retains its $x$- and $y$-coordinates. This staggered set of internal displacements prevents degeneracies in the calculation of $\big(\frac{d\vec{r}}{d\vec{q}}\big)$. More details are reported in the Supplemental Material \cite{supplementalMaterial}).

In order to generate the force field described above, an analytical estimate of $\mathcal{P}(\vec{q})$ must be fit to the samples from the simulation. We model $\mathcal{P}(\vec{q})$ from this data with a Gaussian Mixture Model (GMM) fit using the Expectation-Maximization (EM) algorithm implemented in the Python package Scikit-Learn \cite{pedregosa_scikit-learn_2011-1}.

Figure \ref{fig:pdf_fit} shows the observed histograms for the trimer’s three shape coordinates, overlaid with the corresponding marginals of the GMM fit $\mathcal{P}(\vec{q})$ constructed using 10 components. The close agreement between the GMM and the empirical distributions for each coordinate suggests that 10 Gaussians provide a reasonably accurate representation. This will be further validated in  \ref{sec:Results} through direct comparison of the resulting force field.

\section{Results} \label{sec:Results}

\begin{figure}[t]
\begin{subfigure}{8.6cm}
\includegraphics[width = 8.6cm]
{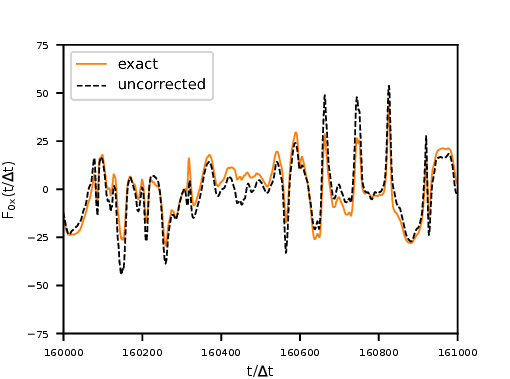}
\end{subfigure}
\begin{subfigure}{8.6cm}
\includegraphics[width = 8.6cm]{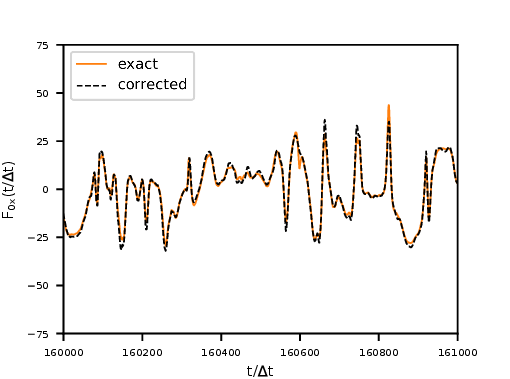}
\end{subfigure}
\caption{\label{fig:forceTimeSeries} Representative segment of the $x$-component force time series acting on a single bead of the trimer. The exact force from simulation (solid line) is compared with the reconstructed force obtained from the gradient of the Gaussian Mixture Model (GMM) fit (dashed line), shown both without (top panel) and with (bottom panel) the rotational entropy correction.}
\end{figure}

\begin{figure}[t]
\begin{subfigure}{8.6cm}
\includegraphics[width = 8.6cm]{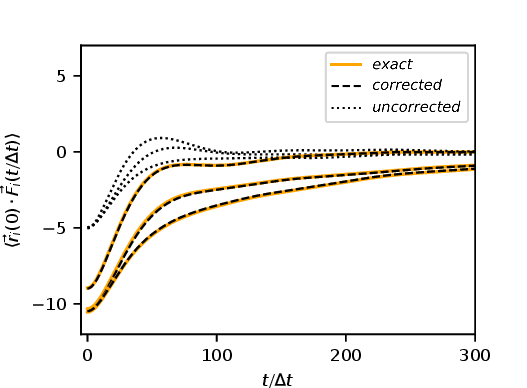}
\end{subfigure}
\begin{subfigure}{8.6cm}
\includegraphics[width = 8.6cm]{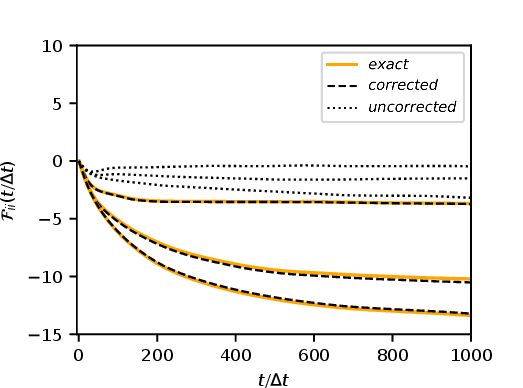}
\end{subfigure}
\caption{\label{fig:tcfComparison} The top panel displays the lagged position–force correlation function for each bead of the model trimer, computed using the exact simulation forces (solid line), the uncorrected gradient of the Gaussian Mixture Model (GMM) fit (dotted line), and the corrected gradient incorporating the rotational entropy term as in Eq.~\ref{eq:force field} (dashed line). The corresponding running integrals $\mathcal{F}_{ii}(t/\Delta t)$ of these correlation functions are shown in the bottom panel using the same line styles.}
\end{figure}


Given a fitted form of $\mathcal{P}(\vec{q})$, obtained, for instance, using a Gaussian Mixture Model, the gradient of the distribution can be evaluated and substituted into Eq. \ref{eq:force field} to compute a time series of generalized forces along the sampled shape trajectories. These generalized forces are then mapped back to the full set of lab frame forces by applying the inverse transformation procedure outlined in Section II of the Supplemental Material \cite{supplementalMaterial}. To isolate the effect of the rotational entropy correction, we compare the force field calculated with and without the second term in Eq. \ref{eq:force field}.

The top panel of Fig. \ref{fig:forceTimeSeries} shows the time series of forces reconstructed from the GMM fit, excluding the rotational entropy correction, plotted against the exact forces from the simulation over a representative segment of the trajectory. Although the overall trends are qualitatively captured, significant discrepancies remain in the finer details of the force profile.

In the bottom panel of Fig.~\ref{fig:forceTimeSeries}, we show the results of the full force reconstruction, now including the rotational entropy correction. The agreement with the exact forces improves markedly, capturing both the overall trends and finer features of the time series.

To evaluate the practical significance of the rotational entropy correction, we turn now to calculate its influence on the reconstruction of time correlation functions (TCFs).

The first TCF we consider is the lagged position–force correlation, which directly contributes to the calculation of the Markovian friction coefficients \cite{frictionHallGuenza}. As such, the accuracy of the reconstructed friction depends critically on the fidelity of the reconstructed force field. In Fig.~\ref{fig:tcfComparison}, we compare observed position–force TCFs for the trimer with the corresponding TCFs computed using the modeled force field, both with and without the rotational entropy correction. Notably, accurate recovery of the TCF is achieved only when the rotational correction is included.

The deviation observed at small $t/\Delta t$ in Fig. \ref{fig:tcfComparison}
underscores the importance of accounting for rotational distortion in the free energy landscape when aiming to accurately capture short-time dynamical behavior. Notably, certain methods for parametrizing friction coefficients or memory kernels in coarse-grained Langevin models rely on the time integral of this correlation function as an input \cite{frictionHallGuenza,yang_numerical_2013}. In the bottom panel of Fig. 3 we see that the error in the total integral of the position-force correlation incurred by ignoring the rotational correction is considerable. As a result, inaccuracies at short times can propagate into the estimation of parameters that govern long-time dynamics. These findings suggest that multiple aspects of CG model construction, based on observed statistics, may be sensitive to the effects of rotational entropy.

To further investigate this possibility, we compute the friction coefficients of the trimer using the Generalized Einstein Relation described in Reference \cite{frictionHallGuenza}. The calculation is performed using three different force fields: the reference force field obtained directly from the simulation trajectory, and two reconstructed force fields derived from the GMM fit to the probability distribution—with and without the inclusion of the rotational entropy correction (see Eq. \ref{eq:force field}).

\begin{figure}[t]
\includegraphics[width = 8.6cm]{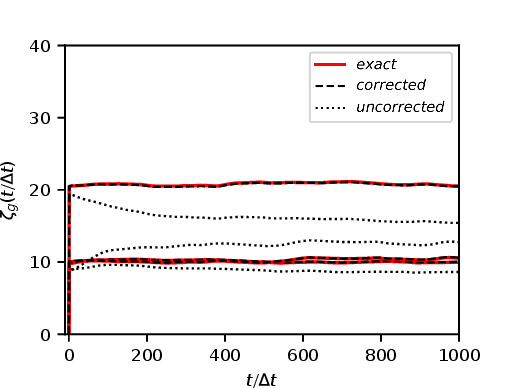}
\caption{\label{fig:friction}
Time-dependent diagonal friction coefficients $\zeta_{g}(t/\Delta t)$ computed using the method of Ref.~\cite{frictionHallGuenza}, based on the exact force field (solid line), the Gaussian Mixture Model (GMM) fit with the rotational entropy correction (dashed line), and the uncorrected GMM fit (dotted line).}
\end{figure}

As shown in Fig.~\ref{fig:friction}, the force field reconstructed from the GMM with the rotational entropy correction reproduces the true friction coefficients
 $\zeta = (10, 10, 20)$. Both exhibit rapid convergence to well-defined plateaus at short times, consistent with the Markovian character of the simulation. In contrast, the friction functions $\zeta_{g}(t)$ obtained from the uncorrected GMM display a slower convergence to their asymptotic limits, indicating artificial memory effects, and ultimately plateau at incorrect values.\\

\section{Discussion}

    Recovering the effective intramolecular force field from atomistic simulations is an essential step in modeling complex macromolecular systems by coarse-graining. In this article, we show that correcting for rotational entropy is, in general, essential for accurately recovering the underlying intramolecular force field from the distribution of internal coordinates in molecules undergoing global translational and rotational diffusion. If left unaccounted for, the rotational-vibrational coupling distorts the apparent free energy landscape (FEL), leading to inaccuracies in the inferred potential of mean force (PMF) and the resulting force field.

    To address this, we derived an explicit correction to the free energy landscape that isolates and removes the rotational entropy contribution, yielding a PMF consistent with the true internal energetics of the molecule. This correction is expressed in terms of the generalized moment of inertia tensor $I^{*}(\vec{q})$, which naturally encodes the dependence of rotational entropy on molecular shape. The corrected PMF can then be used to reconstruct accurate forces and friction coefficients within coarse-grained Langevin dynamics models.

Using a well-controlled toy model—a trimer with analytically specified potential energy, mass, and friction parameters—we validated this approach by comparing the reconstructed force fields and dynamical statistics to known quantities. We showed that inclusion of the rotational entropy correction yields significant improvements in the recovery of the force time series, time correlation functions, and friction coefficients. In contrast, neglecting the correction may lead to observable artifacts, such as spurious memory effects and mis-estimated friction values, even in a simple Markovian setting.

These results underscore that the shape-dependence of rotational entropy is not a negligible artifact but a physically meaningful contribution that can influence both short- and long-time dynamical observables.

By establishing a rigorous formalism and validating it on a tractable model, this work lays the foundation for applying these corrections in more complex settings. In future work, we aim to extend this framework to atomistic molecular dynamics simulations of biomolecules in solution, where the potential of mean force is not known a priori. This will allow us to quantify the practical significance of rotational entropy effects in systems of biological and chemical interest and to assess the conditions under which such corrections are necessary for reliable CG modeling.

\vspace{2mm}

\textit{Acknowledgements-} This material is based upon work supported by the National Science Foundation under Grant No. CHE-2154999. J.M.H. was partially supported as a predoctoral trainee by the NIH-NIGMS Institutional Research Service Award 5T32GM007759-43 in Molecular Biology and Biophysics.
The computational work was partially performed on the
supercomputer Expanse at the San Diego Supercomputer Center, with the support of ACCESS \cite{access} allocation Discover ACCESS CHE100082 (ACCESS is a program supported by the National Science Foundation under Grant No. ACI-1548562). The computational work was partially performed on the University of Oregon high performance computing cluster, Talapas.



\begin{thebibliography}{27}
\bibitem [{Not()}]{Note}%
  \BibitemOpen
  \href@noop {} {\bibinfo  {journal} {At the outset, we note a distinction in terminology: In physics, the force field refers to the field of forces derived as the negative gradient of the potential of mean force, whereas in computational chemistry, the term denotes the mathematical representation of the potential itself. In this manuscript, we follow the physics convention and use force field to refer specifically to the field of forces obtained from the potential of mean force.}\ }\BibitemShut {NoStop}%
\end{thebibliography}

%

\end{document}